\documentclass[draftcls,12pt,onecolumn,oneside]{IEEEtran}
\usepackage{amsmath,amssymb} \interdisplaylinepenalty=2500
\usepackage{color}
\usepackage{subfigure}
\usepackage{epsfig}
\usepackage{graphicx}
\usepackage{dsfont}
\usepackage{verbatim}
\usepackage{amsthm}
\usepackage{setspace}

\usepackage{mathtools}

\usepackage{enumerate}

\newtheorem{definition}{Definition}
\newtheorem{theorem}{Theorem}
\newtheorem{lemma}[theorem]{Lemma}

\newtheorem{claim}[theorem]{Claim}

\newenvironment{remark}{\textit{Remark: }}{}

\newif\ifcomments
\commentstrue 

\newcommand{\calA}{\mathcal{A}}
\newcommand{\calB}{\mathcal{B}}
\newcommand{\calC}{\mathcal{C}}
\newcommand{\calD}{\mathcal{D}}
\newcommand{\calE}{\mathcal{E}}

\newcommand{\calG}{\mathcal{G}}

\newcommand{\calR}{\mathcal{R}}
\newcommand{\calS}{\mathcal{S}}
\newcommand{\calT}{\mathcal{T}}

\newcommand{\calV}{\mathcal{V}}

\newcommand{\calX}{\mathcal{X}}

\newcommand{\bfY}{\mathbf{Y}}

\newcommand{\F}{\mathds{F}}
\newcommand{\Fq}{\mathds{F}_{q}}

\newcommand{\Fbm}{\mathds{F}_{2^{m}}}

\newcommand{\NoE}{E}
\newcommand{\NoS}{S}
\newcommand{\Bl}{n}
\newcommand{\lfs}{m}
\newcommand{\mess}{M}
\newcommand{\Cp}{C}

\newcommand{\Tran}{T}
\newcommand{\TranE}{\hat{T}}

\newcommand{\Basis}{B}
\newcommand{\Ma}{M_1}
\newcommand{\Mb}{M_2}

\newcommand{\Cout}{C_{\mathrm{out}}}
\newcommand{\Cin}{C_{\mathrm{in}}}
\newcommand{\Rout}{R_{\mathrm{out}}}
\newcommand{\Rin}{R_{\mathrm{in}}}
\newcommand{\Ccon}{C_{\mathrm{con}}}
\newcommand{\dout}{d_{\mathrm{out}}}
\newcommand{\din}{d_{\mathrm{in}}}

\newcommand{\bigO}{\ensuremath{\mathcal{O}}}
\newcommand{\smallo}{\ensuremath{o}}

\newcommand{\GVset}{\calA}

\begin{document}

\title{End-to-End Error-Correcting Codes on Networks with Worst-Case Symbol Errors}


\author{{Qiwen Wang} and {Sidharth Jaggi}}

\maketitle
\begin{abstract}
The problem of coding for networks experiencing {\it worst-case symbol errors} is considered. We argue that this is a reasonable model for highly dynamic wireless network transmissions. We demonstrate that in this setup prior network error-correcting schemes can be arbitrarily far from achieving the optimal network throughput. A new {\it transform metric} for errors under the considered model is proposed. Using this metric, we replicate many of the classical results from coding theory. Specifically, we prove new {\it Hamming-type}, {\it Plotkin-type}, and {\it Elias-Bassalygo-type} upper bounds on the network capacity. A commensurate lower bound is shown based on {\it Gilbert-Varshamov-type} codes for error-correction. 
The GV codes used to attain the lower bound can be non-coherent, that is, they do not require prior knowledge of the network topology.
We also propose a computationally-efficient concatenation scheme. 
The rate achieved by our concatenated codes is characterized by a {\it Zyablov-type} lower bound. 
We provide a {\it generalized minimum-distance decoding} algorithm which decodes up to half the minimum distance of the concatenated codes.
The end-to-end nature of our design enables our codes to be overlaid on the classical distributed random linear network codes~\cite{Ho++2003}. Furthermore, the potentially intensive computation at internal nodes for the link-by-link error-correction is un-necessary based on our design.\footnote{This work has been submitted to the IEEE for possible publication. Copyright may be transferred without notice, after which this version may no longer be accessible.}
\end{abstract}

\IEEEpeerreviewmaketitle

\section{Introduction}
A source wishes to transmit information to a receiver over a network with ``noisy" links. Such a communication problem faces several challenges.

The primary challenge we consider is that in highly dynamic environments such as wireless networks, the noise levels on each link might vary significantly across time, and hence be hard to estimate well.
A scenario with the similar variable noise levels challenge is networks with an omniscient adversary. Here the adversary is able to observe all transmissions, and inject into the network errors that depend on the transmitted messages, subject only to a global jamming power constraint.
This issue of variable link noise levels exacerbates at least two other challenges that had been considered settled by prior work.

Firstly, since noise exists in the network, network coding might be dangerous.
This is because all nodes mix information, so even a small
number of symbol errors in transmitted packets may end up corrupting all the information flowing in the network, causing decoding errors. Prior designs for network error-correcting codes exist (for {\it e.g.}~\cite{SonYC:06, Kotter.Kschischang2008}) but as we shall see they are ineffective against symbol errors in a highly dynamic noise setting. In particular, one line of work ({\it e.g.}~\cite{Cai.Yeung2002, Jaggi++2007:Byzantine, Kotter.Kschischang2008,yang2011refined,yang2008weight}) treats even a single symbol error in a packet as corresponding to the entire packet being corrupted, and hence results in rates that are too pessimistic -- the fundamental problem is that the codes are defined over ``large alphabets", and hence are poor at dealing with symbol errors. Another line of work ({\it e.g.}~\cite{SonYC:06}) overlays network coding on link-by-link error correction, but requires accurate foreknowledge of the noise levels on each link to have good performance.

Secondly, in dynamic settings, the coding operations of nodes in the network may be unknown {\it a priori}. Under the symbol error model we consider, 
the ``transform-estimation" strategy of Ho {\it et al.}~\cite{Ho++2003} does not work, since any headers pre-specified 
 can also end up being corrupted.

This work attempts to settle these challenges.
We consider simultaneously the {\it reliability} and {\it universality} issues for random linear network coding. Specifically, we design end-to-end distributed schemes that allow reliable
network communications in the presence of worst-case network noise, wherein the erroneous symbols can be arbitrarily distributed in different network packets with only the constraint that the total number of errors is bounded from above by a certain fraction of all the symbols transmitted in the whole network. With internal network nodes just carrying out linear network coding, error-correction is only accomplished by the receiver(s), who are also able to estimate the linear transform imposed on the source's data by the network.\footnote{As is common in coding theory, the upper and lower bounds on error-correction we prove also directly lead to corresponding bounds on error-detection -- for brevity we omit discussing error-detection in this work.}

As noted above, our codes are robust to a wide variety of network conditions -- whether the symbol errors are evenly distributed among all packets, or even adversarially concentrated among just a few packets, our codes can detect and correct errors up to a network-wide bound on the total number of errors.
Na\"ive implementations of prior codes (for instance, of link-by-link error-correcting codes~\cite{SonYC:06}) that try to correct for worst-case network conditions may result in network codes with much lower rates (see the example in Section~\ref{ex:toy} below). Thus the naturally occurring diversity of network conditions works in our favour rather than against us.

Also, even though our codes correct symbol errors over ``small" finite field rather than errors over larger symbol fields as in prior work, the end-to-end nature of our design enables our codes to be overlaid on classical linear network codes over finite fields, for instance, the random linear network codes of Ho {\it et al}~\cite{Ho++2003}. Further, we free internal nodes from having to implement potentially computationally intensive link-by-link error-correction.
This property might be useful in networks such as sensor networks, where the internal nodes don't have sufficient computational power to perform link-layer error correction.

The main tool used to prove our results is a {\it transform metric} that may be of independent interest (see Section~\ref{sec:metric} for details).
It is structurally similar to the rank-metric used by Silva {\it et al.}~\cite{Silva++2008}, but has important differences that give our codes the power of universal robustness against binary noise, as opposed to the packet-based noise considered in~\cite{Jaggi++2007:Byzantine,Kotter.Kschischang2008,Silva++2008} and~\cite{SonYC:06}.
\vspace{-0.8cm}
\subsection{Previous Work on Network Error Correction}
In general, there are two lines of prior work in the literature on network error control. One approach~\cite{Cai.Yeung2002, Jaggi++2007:Byzantine, Silva++2008} considers correcting packet-level corruptions; the other~\cite{SonYC:06} overlays network coding on link-by-link error correction. 
In 2002, Borade~\cite{borade2002network} proved an information-theoretic outer bound on the rate region for networks with independent and synchronous noisy channels. Simultaneously, Cai and Yeung~\cite{Cai.Yeung2002} considered packet-wise worst-case errors and derived generalized Hamming upper bounds and Gilbert-Varshamov lower bounds for networks.
In 2003, the algebraic network codes of K\"otter and  M{\'e}dard~\cite{Koetter03} and the random linear network codes by Ho {\it et al.}~\cite{Ho++2003} are resilient to node/edge failures that do not change the mincut, which can be considered as packet erasures over large alphabets. 
In 2006, Song, Yeung and Cai~\cite{SonYC:06} proposed a Shannon-type separation theorem for network coding and channel coding in networks consisting of independent channels with random noise, where the channels are not restricted to synchronous ones.
In~\cite{Jaggi++2007:Byzantine}, Jaggi {\it et al.} proposed network codes against adversaries with different attacking power (different numbers of links that the adversaries can eavesdrop/jam). Those schemes are distributed, rate-optimal, and request polynomial design and implementation time.
In~\cite{katti2008symbol}, the authors proposed a layered scheme for improving throughput over wireless mesh networks which has a similar flavor as our work, where the routers (internal nodes) let erroneous bits through without compromising end-to-end reliability.
Silva, K{\"o}tter and Kschischang~\cite{Kotter.Kschischang2008,Silva++2008,silva2011universal,silva2007using,silva2009metrics} used rank-metric codes and subspace codes for networks with packet errors. 
Following the subspace codes by K{\"o}tter and Kschischang, in~\cite{etzion2011error} the authors investigated the coding theory in projective space, and derived bounds corresponding to those by Johnson, Delsarte and Gilvert-Varshamov in the classical coding theory. 
In 2011, we studied the problem on correcting binary errors in random linear network codes in~\cite{wang2011binary}, where we presented the preliminary results including a Hamming-type upper bound and GV-type codes for both coherent and noncoherent networks. 
The works by Yang {\it et al.}~\cite{yang2008weight,yang2011refined} investigated different weight measures for network error correction (with packet errors) and derived counterparts of the Hamming bound, the Singleton bound and the Gilbert-Varshamov bound in the classical coding theory. Although the settings of our work (symbol/bit-level errors) are different from~\cite{yang2008weight,yang2011refined}, the spirit of our work and Yang's work are similar -- refining the bounds in the classical coding theory with novel distance metrics designed for network error correction. 
More recently, Skachek {\it et al.}~\cite{skachek2013hybrid} proposed an extension of subspace codes~\cite{Kotter.Kschischang2008} capable of correcting certain combinations of dimension errors and symbol errors/erasures in noncoherent networks.

\section{Model}\label{sec:model}
\subsection{Network model}\label{sec:model:network}
We model our network by a directed acyclic multigraph\footnote{Our model also allows non-interfering broadcast links in a wireless network to be modeled via a directed hypergraph -- for ease of notation we restrict ourselves to just graphs.}, denoted by $\calG=(\calV,\calE)$, where $\calV$ denotes the set of nodes and $\calE$ denotes the set of edges. A single source node $s\in\calV$ and a set of sinks $\calD\subseteq\calV$ are pre-specified in $\calV$. We denote $|\calE|$ and $|\calD|$, respectively the number of edges and sinks in the network, by $\NoE$ and $\NoS$.
A directed edge $e$ leading from node $u$ to node $v$ can be represented by the vector $(u,v)$, where $u$ is called the {\it tail of $e$} and $v$ is called the {\it head of $e$}. In this case $e$ is called an {\it outgoing edge of $u$} and an {\it incoming edge of $v$}.

The capacity of each edge is one packet -- an length-$\Bl$ vector over a finite field $\Fbm$ -- here
$\Bl$ and $\lfs$ are design parameters to be specified later.
Multiple edges between two nodes are allowed -- this allows us to model links with different capacities.\footnote{By appropriate buffering and splitting edges into multiple edges, any network can be approximated into such a network with unit capacity edges.}
As defined in~\cite{ACLY00}, the {\it network (multicast) capacity}, denoted $\Cp$, is the minimum over all sinks $t \in \calD$ of the mincut of $\calG$ from the source $s$ to the sink $t$. Without loss of generality, we assume there are $\Cp$ edges outgoing from $s$ and incoming edges to $t$ for all sinks $t \in \calD$.
\footnote{In cases where the number of outgoing edges from $s$ (or the number of incoming edges to $t$) is not $\Cp$, we can add a {\it source super-node} (or {\it sink super-node}) with $\Cp$ noiseless {\it edges} connecting to the original source (or sink) of the network. The change in the number of edges and probability of error on each edge are small compared to those of the original network, so our analysis essentially still applies.}
\vspace{-0.5cm}
\subsection{Code model}\label{sec:model:code}
The source node $s$ wants to {\it multicast} a message $\mess$ to each sink $t \in \calD$.
To simplify notation, we consider henceforth just a single sink -- our analysis can be directly extended to the multi-sink case. For brevity we consider bit-flips in this paper. Our results shall translate to general symbol errors with any finite field size. All logarithms in this work are to the base $2$, and we use $H(p)$ to denote the {\it binary entropy function} $-p\log p -(1-p)\log(1-p)$.

\noindent {\bf Random linear network coding:}
All internal nodes in the network perform {\it random linear network coding}~\cite{Ho++2003} over a finite field $\Fbm$. Specifically, each internal node takes uniformly random linear combinations of each incoming packet to generate outgoing packets.
That is, let $e'$ and $e$ index incoming and outgoing edges from a node $v$. The {\it linear coding coefficient from $e'$ to $e$} is denoted by $f_{e',e}\in\Fq$. Let ${\mathbf Y_{e}}$ denote the packet (length-$\Bl$ vector over $\Fbm$) transmitted on the edge $e$. Then $\bfY_{e}=\sum f_{e',e}\bfY_{e'}$, where the summation is over all edges $e'$ incoming to the node $v$, and all arithmetic is performed over the finite field $\Fbm$.

\noindent {\bf Mapping between $\F_2$ and $\Fbm$:}
The noise considered in this work is binary in nature. Hence, to preserve the linear relationships between inputs and outputs of the network,
we use the mappings given in Lemma~$1$ from~\cite{JagEHM:04}. These map addition and multiplication over $\Fbm$ to corresponding (vector/matrix) operations over $\F_2$.
More specifically, a bijection is defined from each symbol (from $\Fbm$) of each packet transmitted on each edge, to a corresponding length-$\lfs$ bit-vector. For ease of notation henceforth, for each edge $e$ and each $i \in \{1,\ldots,\Bl\}$, we use ${\mathbf Y}_e$ and ${\mathbf Y}_e(i)$ solely to denote respectively the length-$\Bl\lfs$ and length-$\lfs$ binary vectors resulting from the bijection operating on packets and their $i$th symbols, rather than the original analogues over $\Fbm$ traversing that edge $e$. Separately, each {\it linear coding coefficient} $f_{e',e} $ at each node is mapped via a homomorphism of abelian groups ($\Fbm$ and $(\F_2)^{m \times m}$) to a specific $\lfs \times \lfs$ binary matrix $F_{e',e} $. The linear mixing at each node is then taken over the binary field -- each length-$\lfs$ binary vector ${\mathbf Y}_{e'}(i)$ (corresponding to the binary mapping of the $i$th symbol of the packet ${\mathbf Y}_{e'}$ over the field $\Fbm$) equals $\sum F_{e',e}{\mathbf Y}_{e'}(i)$. It is shown in~\cite{JagEHM:04} that an isomorphism exists between the binary linear operations defined above, and the original linear network code. In what follows, depending on the context, we use the homomorphism to switch between the scalar (over $\Fbm$) and matrix (over $\F_2$) forms of the network codes' linear coding coefficients, and the isomorphism to switch between the scalar (over $\Fbm$) and vector (over $\F_2$) forms of each symbol in each packet.

In Lemma~\ref{thm:mappinglemma} below, a {\it network coding problem} for a multi-source multi-sink network $\calG$ with source nodes $\calS$ and sinks $\calT$ is defined as an $|\calS| \times |\calT|$ binary matrix, denoted by $\calR_{\calG}=\{ r_{st} \}$, such that $r_{st} = 1$ if the data generated at source $s$ is required at sink $t$, and $0$ otherwise. An $\F_{\beta^m}$-algebraic network code, denoted by $\calC_A(\calG,\beta,m)$, is a network code where the messages on edges and the coding coefficients are elements from a finite field of size $\beta^m$. An $(\F_{\beta})^m$-block network code, denoted by $\calC_B(\calG,\beta,m)$, is a network code where the messages on edges are viewed as vectors of length $m$ over $\F_\beta$ and the coding coefficients are $m \times m$ matrices over $\F_\beta$.
\begin{lemma}\cite{JagEHM:04}[Lemma 1] \label{thm:mappinglemma}
	For any network coding problem $\calR_{\calG}$ solved by an algebraic network code $\calC_A(\calG,p,m)$ there exists an input-output equivalent local reduction to a block network code $\calC_B(\calG,p,m)$.
\end{lemma}

\begin{figure}[!ht] 
    \centering
        \includegraphics[width=0.4\textwidth]{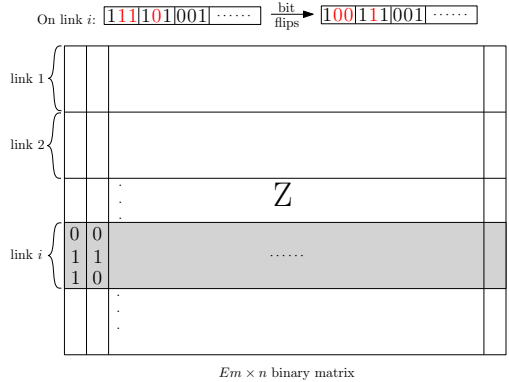}
    \caption{An example of the noise matrix $Z$. The $m \times n$ sub-matrix consisting the $[(i-1)m+1]$th row up to the $im$th row represents the bit-flips on link $i$ in the network.} \label{fig:noisemodel}
\end{figure}

\noindent {\bf Noise:}
We consider {``worst-case noise"} in this work, wherein an arbitrary number of bit-flips can happen in any transmitted packet, subject to the constraint that no more that a fraction of $p$ bits over all transmitted packets are flipped.
The {\it noise matrix} $Z$ is
an $\NoE \lfs\times \Bl$ binary matrix with at most $p\NoE\lfs \Bl$ nonzero entries which can be arbitrarily distributed. In particular, the $\lfs(i-1)+1$ through the $\lfs i$ rows of $Z$ represent the bit flips in the $i$th packet ${\mathbf Y}_{e_i}$ transmitted over the network. If the $(k\lfs + j)$th bit of the length-$\lfs\Bl$ binary vector is flipped, i.e. the $j$th bit of the $k$th symbol over $\Fbm$ in ${\mathbf Y}_{e_i}$ is flipped), then the $(\lfs(i-1)+j,k)$ bit in $Z$ equals $1$, else it equals $0$.
Thus the noise matrix $Z$ represents the {\it noise pattern} of the network. An example of how $Z$ models the bit-flips on the links is shown in Fig.~\ref{fig:noisemodel}.
To model the noise as part of the linear transform imposed by the network, we add an artificial super-node $s'$ connected to all the edges in the network, injecting noise into each packet transmitted on each edge in the network according to entries of the noise matrix $Z$.

\noindent {\bf Source:}
The source has a set of $2^{R\lfs\Bl}$ messages $\{\mess\}$ it wishes to communicate to each sink, where $R$ is the {\it rate} of the source. Corresponding to each message $\mess$ it generates a {\it codeword} $X(\mess)$ using the encoders specified in Section~\ref{sec:achieve:GV:co} (to make notation easier we usually do not explicitly reference the parameter $\mess$ and instead refer simply to $X$). This $X$ is represented by a $\Cp \times \Bl$ matrix over $\Fbm$, or alternatively a $\Cp\lfs \times \Bl$ matrix over $\F_2$. Each row of this matrix corresponds to a packet transmitted over a distinct edge leaving the source.


\noindent {\bf Receiver(s):}
Each sink $t$ receives a batch of $\Cp$ packets. Similarly to the source, it organizes the received packets into a matrix $Y$, which can be equivalently viewed as a $\Cp \times \Bl$ matrix over $\Fbm$ or a $\Cp\lfs \times \Bl$ binary matrix.
Each sink $t$ decodes the message $\hat{\mess}$ from the received matrix $Y$. 

%
\noindent {\bf Transfer matrix and Impulse response matrix:}
Having defined the linear coding coefficients of internal nodes, the packets transmitted on the incoming edges of each sink $t$ can inductively be calculated as linear combinations of the packets on the outgoing edges of $s$. We denote the $\Cp \times \Cp$ {\it transfer matrix from the outgoing edges of $s$ to the incoming edges of $t$ by $\Tran$}, over the finite field $\Fbm$. Alternatively, using the homomorphism described above, $\Tran$ may be viewed as as ${\Cp \lfs \times \Cp \lfs}$ binary matrix.

We similarly define $\TranE$ to be the {\it impulse response matrix}, which is the transfer matrix from an imaginary source $s'$--who injects errors into all edges--to the sink $t$. Note that $\Tran$ is a sub-matrix of $\TranE$, composed specifically of the $\Cp$ columns of $\Tran$ corresponding to the $\Cp$ outgoing edges of $s$.

In this work we require that every $\Cp \times \Cp$ sub-matrix of $\TranE$ is invertible.
As noted in, for instance,~\cite{Ho++2003,Koetter03}
this happens with high probability for random linear network codes. Alternatively, deterministic designs of network error-correcting codes~\cite{Cai.Yeung2002} also have this property.

Using the above definitions the network can thus be abstracted by the equation~\eqref{eq:received} below as a {\it worst-case binary-error network channel}.
\begin{equation}\label{eq:received}
  Y = \Tran X + \TranE Z.
\end{equation}
Similar equations have been considered before (for instance in~\cite{Cai.Yeung2002, Jaggi++2007:Byzantine, Kotter.Kschischang2008}) -- the key difference in this work is that we are interested in $Z$ matrices which are fundamentally best defined over the binary field, and hence, when needed, transform the other matrices in equation~\eqref{eq:received} also into binary matrices.

\noindent{\bf Performance of code:} The source encoders and the decoders at the sinks
together comprise {\it worst-case binary-error-correcting network channel codes}.
A {\it good} worst-case binary-error-correction network channel code has the property that, for all messages $\mess$, and noise patterns $Z$ with at most $p\NoE\lfs\Bl$ bit-flips, the estimated message at the decoder $\hat{\mess} = \mess$.
A rate $R$ is said to be {\it achievable} for the worst-case binary-error channel if, for all sufficiently large $\Bl$, there exists a good code with rate $R$. (As we shall see in Section~\ref{sec:results}, the higher order terms of the achievable rate $R$ is independent of the parameter $m$, {\it i.e.} the alphabet size.)

\section{Motivating Example}~\label{ex:toy}
\begin{figure}[!t] 
    \centering
        \includegraphics[width=0.3\textwidth]{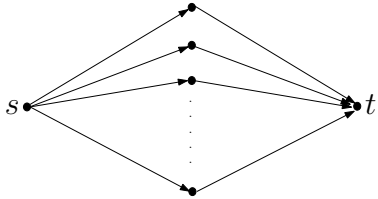}
    \caption{A network with $\Cp$ parallel paths from the source to the destination. Each internal node performs random linear network coding.}     \label{fig:toy}
\end{figure}

We demonstrate via an example that in networks with worst-case bit-errors, prior schemes have inferior performance compared to our scheme. In Figure~\ref{fig:toy}, the network has $\Cp$ paths with a total of $2\Cp$ links that might experience worst-case bit-flip errors ($\Cp \geq 2$).

\noindent {\it Benchmark 1:} If link-by-link error-correction\footnote{Since interior nodes might perform network coding, na\"ive implementations of end-to-end error-correcting codes are not straightforward -- indeed -- that is the primary goal of our constructions.} is applied as in~\cite{SonYC:06}, {\it every} link is then required to be able to correct $2\Cp p m \Bl$ {\it worst-case} bit-flip errors, since all the bit-errors may be concentrated in any single link. Using GV codes (\cite{gilbert1952comparison,varshamov1957estimate}) a rate of $1-H(4 \Cp p)$ is achievable on each link, and hence
the overall rate scales as $\Cp(1-H(4 \Cp p))$. As $\Cp$ increases without bound, the throughput thus actually goes to zero. The primary reason is that every link has to prepare for the worst case number of bit-flips aggregated over the entire network, but in large networks, the total number of bit-flips in the worst-case might be too much for any single link to be able to tolerate.

\noindent {\it Benchmark 2:} Consider now a more sophisticated scheme, combining link-by-link error correction with end-to-end error-correction as in~\cite{Kotter.Kschischang2008}. Suppose each link can correct $\frac{2 \Cp p m \Bl}{k}$ worst-case bit-flips, where $k$ is a parameter to be determined such that the rate is optimized. Then at most $k$ links will fail. Overlaying an end-to-end network error-correcting code as in~\cite{Kotter.Kschischang2008} with link-by-link error-correcting codes such as GV codes (effectively leading to a concatenation-type scheme) leads to an overall rate of $(\Cp-2k)\left(1-H(\frac{4\Cp p}{k})\right)$. For large $\Cp$, this is better than the previous benchmark scheme since interior nodes no longer attempt to correct {\it all} worst-case errors and hence can operate at higher rates -- the end-to-end code corrects the errors on those links that do experience errors. Nonetheless, as we observe below, our scheme still outperforms this scheme, since concatenation-type schemes in general have lower rates than single-layer schemes.

\noindent {\it Our schemes:} The rate achieved by our Gilbert-Varshamov scheme (as demonstrated in Section~\ref{sec:achieve:GV}) is at least $\Cp(1-2H(2p))$. The computationally-efficient concatenated scheme (as demonstrated in Section~\ref{sec:achieve:zyablov}) achieves rate of at least $\max_{0<r<1-2H(2p)} r \cdot \left( 1 - \frac{2p}{H^{-1}\left( \frac{1}{2}(1-r) \right)} \right)$. As can be verified, for small $p$ both our schemes achieve rates higher than either of the benchmark schemes.


\section{Transform Metric} \label{sec:metric}
We first define a ``natural" distance function between binary matrices $\Ma$ and $\Mb$ related as $\Ma=\Mb + \Basis Z$ for a binary {\it basis matrix} $\Basis$ and a binary matrix $Z$.

Let $\Ma$ and $\Mb$ be arbitrary $a \times b$ binary matrices. Let $\Basis $ be a given $a \times c$ matrix with full column rank. Let $\Ma(i)$ and $\Mb(i)$ denote respectively the $i$th columns of $\Ma$ and $\Mb$. We define $d_\Basis(\Ma,\Mb)$, the {\it transform distance between $\Ma$ and $\Mb$ in terms of $\Basis$}, as follows.
\begin{definition}
\label{def:dis}
Let $\delta(i)$ denote the minimal number of columns of $\Basis$ that need to be added to $\Ma(i)$ to obtain $\Mb(i)$. Then the transform distance $d_\Basis(\Ma,\Mb)$ equals $\sum_{i=1}^b \delta(i)$.
\end{definition}

%

\begin{figure}[!t] 
    \centering
        \includegraphics[width=0.45\textwidth]{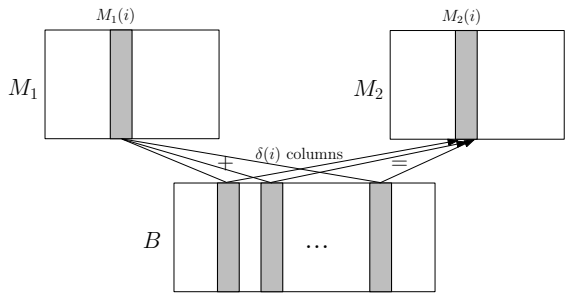}
    \caption{Transform metric: the minimal number of columns of $\Basis$ that need to be added to $\Ma(i)$ to obtain $\Mb(i)$ is $\delta(i)$.} \label{fig:transformmetric}
\end{figure}

The definition of this transform distance is visualized in Figure~\ref{fig:transformmetric}.
The reason we look into this matrix-based metric is because we want to capture how bit-flips in $Z$ perturb $\Tran X$ to the actually received $Y$ (recall in equation~\eqref{eq:received} that $Y = \Tran X + \TranE Z$). In this case, Hamming distance certainly doesn't work, because a very sparse error matrix $Z$ may lead to large Hamming distance between $\Tran X$ and $Y$. In other words, Hamming distance is not able to quantify the noise level of the network.
Notice that the $1$'s in $Z(i)$ choose the corresponding columns of $\TranE$ and add to $\Tran X(i)$. Since any $Cm$ columns from $\TranE$ are linearly independent, if the Hamming weight of $Z(i)$ is less than $Cm$, in the transform distance $\sum_{i=1}^b \delta(i)$, each $\delta(i)$ equals the Hamming weight of $Z(i)$.

\begin{claim} \label{thm:dist}
The function $d_\Basis(\Ma,\Mb)$ is a distance measure.
\end{claim}

\begin{remark}
	It can be proved directly that $d_\Basis(\Ma,\Mb)$ is a metric. A more conceptual proof relates to coset decoding and Cayley graph.
	
	Firstly, consider the case when $n=1$, {\it i.e.}, when the matrices $\Ma$ and $\Mb$ reduce to column vectors. Let $B$ be a binary $a \times c$ matrix with full column rank, which hence implies $a \leq c$. We can regard $B$ as a parity-check matrix of a binary code $\calC$ of length $c$ and dimension $c-a$. the function $d_\Basis(\Ma,\Mb)$ is closely related to coset decoding of $\calC$. A column vector $S$ of length $a$ can be interpreted as the syndrome vector. In coset decoding, we want to find the minimum number of columns in $B$ which sum to the vector $S$. This is called the syndrome weight in~\cite{jurrius2009extended} or the coset leader weight in~\cite{helleseth1979weight}. Hence, $\delta(i)$ defined in Definition~\ref{def:dis} is the same as the syndrome weight of $\Ma -\Mb$ with respect to the parity-check matrix $B$. The fact that the syndrome weight induces a metric on $\F_2^a$ can be seen by considering the Cayley graph on $\F_2^a$, generated by the columns of $B$, {\it i.e.}, the vertices of the Cayley graph are identified with the vectors in $\F_2^a$, and two vertices are adjacent if and only if their difference is one of the columns in $B$. Then, $d_\Basis(\Ma,\Mb)$ is the length of a shortest path from $\Ma$ to $\Mb$ in the graph. This is the graph distance and is hence a metric satisfying the triangle inequality. For $n \leq 2$, $d_\Basis(\Ma,\Mb)$  is the sum of $n$ syndrome weights, and therefore is also a metric.\footnote{This relation between our transform metric and the syndrome weight/coset leader weight in cose decoding is pointed out by Prof. Kenneth W. Shum.}
\end{remark}

\section{Main Results} \label{sec:results}
%
%
%
By using the transform metric, we derive converses and design achievable schemes which can be viewed as counterparts of classical coding theory. In this section, we present our main results.

\begin{theorem}[Hamming-Type Bound]
\label{thm:Hammingbd1} For all $p$ less than $\frac{\Cp}{2\NoE\lfs}$, an upper bound on the achievable rate of any code over the worst-case binary-error channel is $1-\frac{\NoE}{\Cp}H(p)+ \smallo\left(\frac{\log(\NoE\lfs n+1)}{\Cp \lfs n }\right)$.
\end{theorem}


\begin{theorem}[Plotkin-Type Bound]
\label{thm:PlotkinMain}
\mbox{}
\begin{enumerate}
\item For networks with $E \geq 2C$, 
	\begin{enumerate}[i.]
	\item for all $p$ less than $\left( 1 - \frac{C}{E} \right) \frac{C}{E}$, an upper bound on the achievable rate of any code over the worst-case binary-error network channel is $1 - \frac{E^2}{CE-C^2} p$;
	\item if $p$ is greater than $\left( 1 - \frac{C}{E} \right) \frac{C}{E}$, asymptotically the rate achieved by any code over the worst-case binary-error network channel is $0$.
	\end{enumerate}
\item For networks with $E < 2C $,
	\begin{enumerate}[i.]
	\item for all $p\leq 1/4$,  an upper bound on the achievable rate of any code over the worst-case binary-error network channel is $1-4p$;
	\item if $p > 1/4$,  asymptotically the rate achieved by any code over the worst-case binary-error network channel is $0$.
	\end{enumerate}
\end{enumerate}
\end{theorem}

\begin{theorem}[Elias-Bassalygo-Type Bound]
\label{thm:Elias}
For all $p$ less than $\frac{C}{2Em} \left( 1 - \frac{C}{2Em}  \right)$, an upper bound on the achievable rate of any code over the worst-case binary-error network channel is $1 - \frac{E}{C} H\left(\frac{1-\sqrt{1-4p}}{2}\right)+ \smallo\left(\frac{\log(Emn+1)}{Cmn}\right)$.
\end{theorem}

%

\begin{theorem}[Gilbert-Varshamov-Type Bound] \label{thm:GV}
\mbox{}
\begin{enumerate}
\item Coherent GV-type network codes achieve a rate of at least $1-\frac{\NoE}{\Cp}H(2p)- \smallo\left(\frac{\log(2p\NoE \lfs \Bl+1)}{\Bl}\right)$.\label{thm:GVbdco} 
\item Non-coherent GV-type network codes achieve a rate of at least $1-\frac{\NoE}{\Cp}H(2p)- \smallo\left( \frac{\log(2p\NoE \lfs \Bl+1)+\NoE}{\Bl} \right)$.\label{thm:GVbdInco} 
\end{enumerate}
\end{theorem}


\begin{theorem}[Zyablov-Type Bound]
\label{thm:zyablovbd} 
	Concatenation network codes achieve a rate of at least 
\begin{equation}
\max_{0<r<1-\frac{E}{C}H(2p)} r \cdot \left( 1 - \frac{2p}{H^{-1}\left( \frac{C}{E}(1-r) \right)} \right). \nonumber
\end{equation}	
\end{theorem}

\begin{figure} [!t] 
    \centering
        \includegraphics[width=0.45\textwidth]{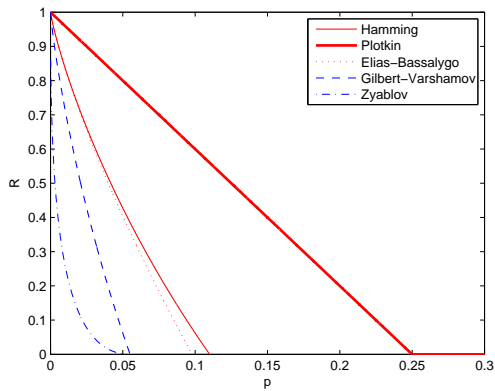}
    \caption{The plot of our results in the senario for a diamond network where $E=8$, $C=4$, and $m=2$.} \label{fig:plotbounds01}
\end{figure}

\begin{figure} [!t] 
    \centering
        \includegraphics[width=0.45\textwidth]{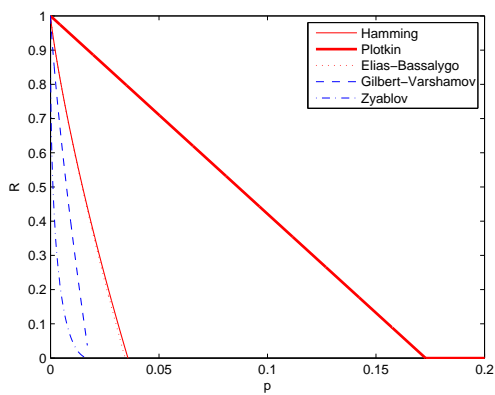}
    \caption{The plot of our results in the senario for the butterfly network where $E=9$, $C=2$, and $m=3$.} \label{fig:plotbounds02}
\end{figure}

In Fig.~\ref{fig:plotbounds01} and Fig.~\ref{fig:plotbounds02} we plot our upper and lower bounds for two different scenarios.

\begin{remark}
If we set $C=E=m=1$, {\it i.e.,} the classical point-to-point worst-case binary-error channel, all our bounds in Theorems~\ref{thm:Hammingbd1}-\ref{thm:zyablovbd} reduce to classical Hamming bound, Plotkin bound, Elias-Bassalygo bound, Gilbert-Varshamov bound, and Zyablov bound.
\end{remark}

\section{Converse} \label{sec:converse}
In this section, we prove lower bounds (Theorems~\ref{thm:Hammingbd1},~\ref{thm:PlotkinMain} and~\ref{thm:Elias}) on information rates that can be communicated by any code through networks with worst-case bit-flips. 
In Section~\ref{sec:converse:Hamming}, we derive Hamming-type upper bounds. Our bounding technique is motivated by the corresponding Hamming bound technique in classical coding theory~\cite{Hamming:50} -- the main challenge lies in deriving good {\it lower} bounds for the ``volumes of spheres" in our network channel model and corresponding transform metric defined in Section~\ref{sec:metric}.
In Section~\ref{sec:converse:Plotkin}, a Plotkin-type upper bound as stated in Theorem~\ref{thm:PlotkinMain} is proved, in which a constraint on the fraction of error $p$ for achieving positive rates is derived.
Section~\ref{sec:converse:elias} proves the Elias-Bassalygo-type upper bound as stated in Theorem~\ref{thm:Elias}, which is a tighter bound than the Hamming-type bound for $p$ less than $\frac{C}{2Em} \left( 1 - \frac{C}{2Em}  \right)$.
\vspace{-0.5cm}
\subsection{Hamming-Type Bound}\label{sec:converse:Hamming}
Since each transmitted codeword $X$ is a $\Cp \lfs \times \Bl$ binary matrix, the number of possible choices of $X$ is at most $2^{\Cp \lfs \Bl}$.
Suppose $X$ is transmitted, by the definitions of the worst-case bit-error network channel~\eqref{eq:received}, the received $Y$ lies in the radius-$p\NoE\lfs\Bl$ ball (in the transform metric)
$\calB_{\TranE}(\Tran X,p\NoE\lfs\Bl)$ defined as 
	$\calB_{\TranE}(\Tran X,p\NoE\lfs\Bl) = \{Y|d_{\TranE}(\Tran X,Y)\leq p\NoE\lfs\Bl\}.$
For the message corresponding to $X$ to be uniquely decodable, it is necessary that the balls $\calB_{\TranE}(\Tran X,p\NoE\lfs\Bl)$ be non-intersecting for each $X$ chosen to be in the codebook.
Hence to get an upper bound on the number of codewords that can be chosen, we need to derive a lower bound of the volume of $\calB_{\TranE}(\Tran X,p\NoE\lfs\Bl)$. Recall that $Y$ equals $\Tran X+\TranE Z$. Hence we need to bound from below the number of distinct values of $\TranE Z$ for $Z$ with at most  $p\NoE \lfs \Bl$ $1$'s.

We consider the case that all $Z$'s has exactly $p\NoE \lfs n $ $1$'s.
We now show that, in this case, every such distinct matrix $Z$ results in distinct $\TranE Z$.
Suppose not -- in that case there exist distinct $Z$ and $Z'$ with $p\NoE \lfs n$ $1$'s in both matrices such that $\TranE Z$ equals $\TranE Z'$, {\it i.e.}, $\TranE (Z-Z')$ equals the zero matrix. In particular, for the $i$th column of $Z$ and $Z'$, denoted by $Z(i)$ and $Z'(i)$, it must be the case that $\TranE \left( Z(i) - Z'(i) \right)$ equals $0$. 
We consider the number of $1$'s in $Z(i)-Z'(i)$ and denote it by $\theta_i$.

We now view $\TranE$ and $Z$ as matrices over $\Fbm$. 
As to the matrix $\TranE$ viewed over $\Fbm$, since we are deriving a worst-case upper bound, we require that every $\Cp \times \Cp$ sub-matrix of $\TranE$ is invertible (as noted before 
this happens with high probability for random linear network codes). Hence $\TranE (Z(i)-Z'(i))$ equals the zero vector means that the number of nonzero elements in $Z(i)-Z'(i)$ over $\Fbm$ is at least $C$, because any linear combination of less than $C$ columns from $\TranE$ should be a nonzero vector. Hence, the number of $1$'s in $Z(i)-Z'(i)$ over $\F_2$ is also at least $C$ (since an element over $\Fbm$ is zero if and only if each of the $\lfs$ bits in its binary representation is zero). Sum up over all the columns, we have $\sum_{i=1}^{n} \theta_i \geq Cn > 2pEmn$, which contradicts with the assumption that all $Z$'s has exactly $p\NoE \lfs n $ $1$'s.


Hence the number of distinct values for
$\TranE Z$ is at least the number of distinct values for $Z$ with $p\NoE \lfs n$ $1$'s. This equals is at least ${{\NoE\lfs n}\choose{p\NoE\lfs n}}$, which by Stirling's approximation~\cite{CoverT:91} gives us that 
\begin{equation} \label{equ:hammingtypestirling}
 |\calB_{\TranE}(\Tran X,p\NoE\lfs\Bl)| \geq 2^{\NoE\lfs n H(p)-\log(\NoE\lfs n+1)}. 
\end{equation}

The total number of $\Cp \lfs \times \Bl$ binary matrices is $2^{\Cp \lfs \Bl}$.
Thus an upper bound on the size of any codebook for the worst-case binary-error channel is
\begin{equation*}
    \frac{2^{\Cp \lfs \Bl}}{2^{\NoE \lfs \Bl H(p)-\log(\NoE\lfs n +1)}} = 2^{ \left( 1-\frac{\NoE}{\Cp}H(p)+\frac{\log(\NoE\lfs n +1)}{\Cp \lfs \Bl} \right) \Cp \lfs \Bl},
\end{equation*}
which, asymptotically in $\Bl$, gives the Hamming-type upper bound on the rate of any code as $1-\frac{\NoE}{\Cp}H(p) + \smallo\left( \frac{\log(\NoE\lfs n+1)}{\Cp \lfs \Bl}\right)$.

\subsection{Plotkin-Type Bound} \label{sec:converse:Plotkin}
In this section, we derive a Plotkin-type upper bound on the achievable rate over worst-case bit-flip networks. 
The argument is in the fashion of the conventional Plotkin bound~\cite{plotkin1960binary}.
In Lemma~\ref{thm:PlotkinA} and Lemma~\ref{thm:PlotkinB} below, we first bound the codebook sizes for different parameter regimes of the minimum distance (in transform metric).

\begin{definition}
\label{def:transcodebook}
	Let $\calX \subseteq \{ {\F}_2^{Cm \times n} \}$ be a codebook for the worst-case binary-error network channel, the transformed codebook $T \calX$ is obtained by multiplying every codeword in $\calX$ by $T$.
\end{definition}

\begin{lemma}
\label{thm:PlotkinA}
Let $\calX \subseteq \{ {\F}_2^{Cm \times n} \}$ be a codebook for the worst-case binary-error network channel with block length $n$, and its transformed codebook $T \calX$ with minimum transform distance $d$.
\begin{enumerate}
\item For networks with $E \geq 2C$, if $d > 2\left(1-\frac{C}{E}\right)Cmn$, then $|\calX| \leq \frac{d}{d - 2\left(1-\frac{C}{E}\right)Cmn}$,
\item For networks with $ E < 2C$, if $d > Emn/2$, we have $|\calX|  \leq \frac{2d}{2d - Emn}$.
\end{enumerate}
\end{lemma}
\noindent {\it Proof:}
Denote the codebook size by $M$, and let $X_1, X_2, \dots X_M$ be the $M$ codewords in $\calX$. Then we have $d_{\TranE}(\Tran X_i, \Tran X_j) \geq d$ for all $i \neq j$.
Hence we can bound the sum of all distances 
\begin{equation} 
	\sum\nolimits_{1 \leq i \leq j \leq M} d_{\TranE}(\Tran X_i, \Tran X_j) \geq {M \choose 2} d. \label{equ:sumdistI}
\end{equation}
On the other hand, considering the columns of the codewords, we have from the definition $d_{\TranE}(\Tran X_i, \Tran X_j) = \sum\nolimits_{k=1}^{n} \delta_{ij}(k)$, where $\delta_{ij}(k)$ is the minimum number of columns from $\TranE$ that need to be added to $\Tran X_i(k)$ to obtain $\Tran X_j(k)$. Hence,
\begin{align}
	\sum\nolimits_{1 \leq i \leq j \leq M} d_{\TranE}(\Tran X_i, \Tran X_j) 
	& =  \sum\nolimits_{1 \leq i \leq j \leq M} \sum\nolimits_{k=1}^{n} \delta_{ij}(k)  \nonumber \\ 
	& = \sum\nolimits_{k=1}^{n} \sum\nolimits_{1 \leq i \leq j \leq M} \delta_{ij}(k). \label{equ:sumdistcolumn}
\end{align}
Now we focus on an arbitrary column $k$, and we want to characterize an upper bound on the sum of transform metric distances of the set of column vectors $\{ \Tran X_1(k), \Tran X_2(k), \dots, \Tran X_M(k)    \}$.
We require $\TranE$ to have full column rank (as noted before this happens with high probability for random linear network codes), we can write $\Tran X_i(k) =  \TranE V_i$ for some binary vector $V_i$ for all $1\leq 1\leq M$. Moreover, we require the Hamming weight of $V_i$ to be no more than $Cm$, because we choose the least number of columns from $\TranE$ which sum up to $\Tran X_i(k)$ and the column rank of $\TranE$ equals $Cm$ as we required. Hence $\delta_{ij}(k)$ can be bounded from above by the Hamming distance between $V_i$ and $V_j$. So we have
\begin{equation}
	\sum\nolimits_{1 \leq i \leq j \leq M} \delta_{ij}(k) \leq \sum\nolimits_{1 \leq i \leq j \leq M} d_{\mathrm{Hamming}}(V_i,V_j). \label{equ:sumHamming}
\end{equation}
Now we use a similar argument as for the proof of the conventional Plotkin bound. We arrange the vectors $V_1, V_2, \dots, V_M$ to an $M \times Em$ binary matrix, where the $i$th row of the matrix correspond to vector $V_i$. Suppose for column $l$ of the matrix, there are $s_l$ $1$'s and $M - s_l$ $0$'s. Then
\begin{equation}
	\sum\nolimits_{1 \leq i \leq j \leq M} d_{\mathrm{Hamming}}(V_i,V_j) = \sum\nolimits_{l=1}^{Em} s_l (M - s_l). \label{equ:sumsl}
\end{equation}

We differentiate between two cases according to the network parameters $C$ and $E$.

\noindent {\bf Case 1 ($E \geq 2C$):}
In this case, we can't let $s_l = M/2$. Otherwise the $V_i$'s have average Hamming weight $Em/2$, which is larger than the constraint $Cm$.
Recall that we have the constraint that $W_{\mathrm{Hamming}}(V_i) \leq Cm$. Hence the total number of $1$'s in the matrix is bounded from above by $ \sum\nolimits_{l=1}^{Em} s_l  \leq MCm$. Hence,
\begin{align}
 \sum\nolimits_{l=1}^{Em} s_l (M - s_l)
 & = M \sum\nolimits_{l=1}^{Em} s_l - \sum\nolimits_{l=1}^{Em} {s_l}^2 \nonumber \\
 & \leq M \sum\nolimits_{l=1}^{Em} s_l - \left( \sum\nolimits_{l=1}^{Em} s_l \right)^2 / {Em} \nonumber \\
 & \leq M^2Cm - \frac{(MCm)^2}{Em} \nonumber \\
 & = M^2 \left(1-\frac{C}{E} \right)Cm \label{equ:sumdistbd}
\end{align}
Combining equations~\eqref{equ:sumdistcolumn}, ~\eqref{equ:sumHamming}, ~\eqref{equ:sumsl} and~\eqref{equ:sumdistbd}, we have
\begin{equation}
 \sum\nolimits_{1 \leq i \leq j \leq M} d_{\TranE}(\Tran X_i, \Tran X_j) \leq n M^2 \left(1-\frac{C}{E}\right)Cm. \label{equ:sumdistII}
\end{equation}
From equations~\eqref{equ:sumdistI} and~\eqref{equ:sumdistII}, we have
\begin{equation}
	 M(M-1)d/2 \leq n M^2\left(1-\frac{C}{E}\right)Cm.  \nonumber
\end{equation}
Hence when $d > 2\left(1-\frac{C}{E}\right)Cmn$, we have $|\calX| = M \leq \frac{d}{d - 2\left(1-\frac{C}{E}\right)Cmn}$.

\noindent {\bf Case 2 ($E < 2C$):}
In this case, let $s_l = M/2$ to maximize equation~\eqref{equ:sumsl}. Hence,
 \begin{align}
 \sum\nolimits_{l=1}^{Em} s_l (M - s_l)
 & \leq   \sum\nolimits_{l=1}^{Em} M^2/4 \nonumber \\
 & = EmM^2/4.  \label{equ:sumdistbd2}
\end{align}
Combining equations~\eqref{equ:sumdistcolumn}, ~\eqref{equ:sumHamming}, ~\eqref{equ:sumsl} and~\eqref{equ:sumdistbd2}, we have
\begin{equation}
 \sum\nolimits_{1 \leq i \leq j \leq M} d_{\TranE}(\Tran X_i, \Tran X_j) \leq nEmM^2/4. \label{equ:sumdistII2}
\end{equation}
From equations~\eqref{equ:sumdistI} and~\eqref{equ:sumdistII2}, we have
\begin{equation}
	 M(M-1)d/2 \leq n EmM^2/4.  \nonumber
\end{equation}
Hence when $d > Emn/2$, we have $|\calX| = M \leq \frac{2d}{2d - Emn}$.
\hfill $\Box$

\begin{lemma}
\label{thm:PlotkinB}
Let $\calX \subseteq \{ {\F}_2^{Cm \times n} \}$ be a codebook for the worst-case binary-error network channel with block length $n$, and its transformed codebook $T \calX$ with minimum transform distance $d$. 
\begin{enumerate}
\item For networks with $E \geq 2C$, if $d \leq 2\left(1-\frac{C}{E}\right)Cmn$, then 
\begin{equation}
|\calX| \leq d \cdot 2^{Cmn-\frac{E}{2(E-C)}d+\frac{E}{2(E-C)}}, \nonumber
\end{equation}
\item For networks with $E < 2C $, if $d \leq Emn/2$, we have 
\begin{equation}
|\calX| \leq  2d \cdot 2^{Cmn - \frac{2C}{E}(d-1)}. \nonumber
\end{equation}
\end{enumerate}
\end{lemma}
\noindent {\it Proof:}
\noindent {\bf Case 1 ($E \geq 2C$):}
When $d \leq 2 \left(1-\frac{C}{E}\right)Cmn$, we have $n \geq \frac{E}{2C(E-C)m}d$.
Let $l = n - \frac{E}{2C(E-C)m}(d-1)$, for each matrix $G \in \F_2^{Cm \times l}$, let $\calX_G$ be a subcode of $\calX$ consisting of all codewords which have $G$ as the $Cm \times l$ submatrix in the first $l$ columns, then puncture the first $l$ columns. Formally,
\begin{equation}
	\calX_G = \{  X^{[l+1,n]}  | X(i) = G(i) \mbox{ for }  1 \leq i \leq l  \}, \nonumber
\end{equation}
where $X^{[l+1,n]} \in \F_2^{Cm \times (n-l)}$ is the submatrix of $X$ consisting of the $n-l$ columns $X(l+1), X(l+2), \dots, X(n)$.
For each $G$, the subcode $\calX_G$ is a codebook with block length $n-l = \frac{E}{2C(E-C)m}(d - 1)$.
The original codebook $\calX$ has minimum transform metric distance $d$, so does the subcode $\calX_G$. 
Hence $d > 2(1-\frac{C}{E})Cm \left( \frac{E}{2C(E-C)m}(d - 1) \right) = d - 1$, and by Lemma~\ref{thm:PlotkinA}.1 we have $|\calX_G| \leq d$. The original codebook size can relate to the sizes of the subcodes as $|\calX| = \sum\nolimits_{G \in \F_2^{Cm \times l}} |\calX_G|$. Hence we have $|\calX| \leq d \cdot 2^{Cml} = d \cdot 2^{Cm\left(n-\frac{E}{2C(E-C)m}d+\frac{E}{2C(E-C)m}\right)}$.

\noindent {\bf Case 2 ($E < 2C$):}
When $d \leq Emn/2$, we have $n \geq 2d/{Em}$. Let $l = n - \frac{2}{Em}(d-1)$, for each matrix $G \in \F_2^{Cm \times l}$, let $\calX_G$ be a subcode of $\calX$ consisting of all codewords which have $G$ as the $Cm \times l$ submatrix in the first $l$ columns, then puncture the first $l$ columns. Formally,
\begin{equation}
	\calX_G = \{  X^{[l+1,n]}  | X(i) = G(i) \mbox{ for }  1 \leq i \leq l  \}, \nonumber
\end{equation}
where $X^{[l+1,n]} \in \F_2^{Cm \times (n-l)}$ is the submatrix of $X$ consisting of the $n-l$ columns $X(l+1), X(l+2), \dots, X(n)$.
For each $G$, the subcode $\calX_G$ is a codebook with block length $n-l = \frac{2}{Em}(d-1)$.
The original codebook $\calX$ has minimum transform metric distance $d$, hence so does the subcode $\calX_G$. 
Hence $d > Em/2 \cdot \frac{2}{Em}(d-1) = d - 1$, and by Lemma~\ref{thm:PlotkinA}.2 we have $|\calX_G| \leq 2d$. The original codebook size can be related to the sizes of the subcodes as $|\calX| = \sum\nolimits_{G \in \F_2^{Cm \times l}} |\calX_G|$. Hence we have $|\calX| \leq 2d \cdot 2^{Cml} = 2d \cdot 2^{Cmn - \frac{2C}{E}(d-1)}$.
\hfill $\Box$

With Lemma~\ref{thm:PlotkinA} and Lemma~\ref{thm:PlotkinB}, our Plotkin-type upper bound on the asymptotic optimal rate follows naturally.

Any codebook needs to have minimum transform metric distance at least $d = 2pEmn+1$. 

\noindent {\bf Case 1 ($E \geq 2C$):}
If $p > \left( 1 - \frac{C}{E} \right) \frac{C}{E}$, the minimum distance $d = 2pEmn+1 > 2\left(1-\frac{C}{E}\right)Cmn$ and by Lemma~\ref{thm:PlotkinA}.1, the codebook size is of order $\bigO(n)$, because we can let $d = 2\left(1-\frac{C}{E}\right)Cmn + \alpha$ for some constant $\alpha$, then $|\calX| \leq d/\alpha$, which is linear in $d$ hence also linear in $n$. Hence the rate goes to $0$ as $n \to \infty$.

When $p \leq \left( 1 - \frac{C}{E} \right) \frac{C}{E}$, the minimum distance $d = 2pEmn+1 \leq 2\left(1-\frac{C}{E}\right)Cmn$ and by Lemma~\ref{thm:PlotkinB}, the size of any codebook is bounded from above by $ |\calX| \leq d \cdot 2^{Cm \left( n-\frac{E}{2C(E-C)m}d+\frac{E}{2C(E-C)m} \right) }$. Hence, asymptotically
\begin{align*}
	R 
	& = \lim_{n \to \infty} \frac{1}{Cmn} \log{|\calX|} \\
	& \leq \lim_{n \to \infty} \frac{1}{Cmn} \left( \log{d} + Cmn-  \frac{E}{2(E-C)}d +   \frac{E}{2(E-C)} \right) \\
	& = \lim_{n \to \infty} \frac{1}{Cmn} \left( \log{d} + Cmn-  \frac{E}{2(E-C)}(2pEmn+1) +   \frac{E}{2(E-C)} \right) \\
	& = \lim_{n \to \infty} \frac{1}{Cmn} \left( \log{d} + Cmn-  \frac{E^2Mn}{E-C} p \right) \\
	& = 1-\frac{E^2}{CE-C^2} p.
\end{align*}

\noindent {\bf Case 2 ($E < 2C$):}
If $p > 1/4$, the minimum distance $d = 2pEmn+1 > Emn/2$ and by Lemma~\ref{thm:PlotkinA}.2, the codebook size is of order $\bigO(n)$ as pointed out above. Hence the rate goes to $0$ as $n \to \infty$.

When $p \leq 1/4$, the minimum distance $d = 2pEmn+1 \leq Emn/2$ and by Lemma~\ref{thm:PlotkinB}.2, the size of any codebook is bounded from above by $|\calX| \leq 2d \cdot 2^{Cml} = 2d \cdot 2^{Cmn - \frac{2C}{E}(d-1)}$.
Hence, asymptotically
\begin{align*}
	R 
	& = \lim_{n \to \infty} \frac{1}{Cmn} \log{|\calX|} \\
	& \leq \lim_{n \to \infty} \frac{1}{Cmn} \left( \log{2d} + Cmn -  \frac{2C}{E}(d-1) \right) \\
	& = \lim_{n \to \infty} \frac{1}{Cmn} \left( \log{2d} + Cmn-  \frac{2C}{E}2pEmn \right) \\
	& = \lim_{n \to \infty} \frac{1}{Cmn} \left( \log{2d} + Cmn-  4Cmnp \right) \\
	& = 1-4p.
\end{align*}

\subsection{Elias-Bassalygo-Type Bound} \label{sec:converse:elias}
In this section, we derive an Elias-Bassalygo-type upper bound on achievable rates over worst-case bit-flip networks. The argument is in the fashion of the conventional Elias-Bassalygo bound~\cite{bassalygo1965new}.
Firstly, a Johnson-type bound based on our transform metric is proved in the following Lemma~\ref{thm:Johnson}. 
\begin{lemma}[Johnson-Type Bound]
\label{thm:Johnson}
Let $J_{\TranE}(Cm \times n, d, e)$ be the maximum number of codewords in a ball of transform metric radius $e$ for any transformed codebook by matrix $\Tran$ with minimum transform metric distance $d$. If $\frac{e}{Emn} < \frac{1}{2}\left( 1 - \sqrt{1-\frac{2d}{Emn}}  \right) $, then 
\begin{equation}
J_{\TranE}(Cm \times n, d, e) \leq \frac{dEmn}{2}. \nonumber
\end{equation}
\end{lemma}
\noindent {\it Proof:}
For a transformed codebook $T \calX$ with minimum transform metric distance $d$, and a matrix $O \in \F_2^{Cm \times n}$ being the center of the ball $\calB_{\TranE}(O,e)$.
Let $X_1, X_2, \dots, X_M$ be the codewords in $\calX$ where $M$ denotes the codebook size. Define $X'_i = X_i - O$ for all $1 \leq i \leq M$, that is, consider the ball $\calB_{\TranE}(O,e)$ and the codebook shifted to the zero matrix $0^{Cm \times n}$.
Then for the shifted codewords $X'_i$'s: 1) For $1\leq i\leq M$, $d_{\TranE}(X'_i,0^{Cm \times n}) \leq e$; 2) For $i \neq j$, $d_{\TranE}(TX'_i,TX'_j) \geq d$.

The following proof is almost the same as that of Lemma~\ref{thm:PlotkinA}, except with the additional weight constraint $d_{\TranE}(X'_i,0^{Cm \times n}) \leq e$.

The sum of distances of the shifted code book can be bounded from below by
\begin{equation}
	\sum\nolimits_{1 \leq i \leq j \leq M} d_{\TranE}(\Tran X'_i, \Tran X'_j) \geq {M \choose 2} d. \label{equ:JohnsumdistI}
\end{equation}
On the other hand, considering the columns of the shifted codewords, we have from the definition $d_{\TranE}(\Tran X'_i, \Tran X'_j) = \sum\nolimits_{k=1}^{n} \delta'_{ij}(k)$, where $\delta'_{ij}(k)$ is the minimum number of columns from $\TranE$ that need to be added to $\Tran X'_i(k)$ to obtain $\Tran X'_j(k)$. Hence,
\begin{align}
	\sum\nolimits_{1 \leq i \leq j \leq M} d_{\TranE}(\Tran X'_i, \Tran X'_j) 
	& =  \sum\nolimits_{1 \leq i \leq j \leq M} \sum\nolimits_{k=1}^{n} \delta'_{ij}(k)  \nonumber \\
	& = \sum\nolimits_{k=1}^{n} \sum\nolimits_{1 \leq i \leq j \leq M} \delta'_{ij}(k). \label{equ:Johnsumdistcolumn}
\end{align}
Now we just focus on some column $k$, and we want to characterize an upper bound on the sum of transform metric distances of the set of column vectors $\{ \Tran X'_1(k), \Tran X'_2(k), \dots, \Tran X'_M(k) \}$.
We require $\TranE$ to have full column rank (as noted before this happens with high probability for random linear network codes), we can write $\Tran X'_i(k) =  \TranE V'_i$ for some binary vector $V'_i$ for all $1\leq 1\leq M$. Moreover we choose the least number of columns from $\TranE$ which sum up to $\Tran X'_i(k)$, that is, $V'_i$ has the least possible Hamming weight. Hence $\delta'_{ij}(k)$ can be bounded from above by the Hamming distance between $V'_i$ and $V'_j$. So we have
\begin{equation}
	\sum\nolimits_{1 \leq i \leq j \leq M} \delta'_{ij}(k) \leq \sum\nolimits_{1 \leq i \leq j \leq M} d_{\mathrm{Hamming}}(V'_i,V'_j). \label{equ:JohnsumHamming}
\end{equation}
Arranging vectors $V'_1, V'_2, \dots, V'_M$ to an $M \times Em$ binary matrix, where the $i$th row correspond to vector $V'_i$. Suppose for column $l$ of the matrix, there are $s'_l$ $1$'s and $M - s'_l$ $0$'s. Then
\begin{equation}
	\sum\nolimits_{1 \leq i \leq j \leq M} d_{\mathrm{Hamming}}(V'_i,V'_j) = \sum\nolimits_{l=1}^{Em} s'_l (M - s'_l). \label{equ:Johnsumsl}
\end{equation}
Let $e'_k = \frac{\sum\nolimits_{l=1}^{Em} S'_l}{M}$, that is, the average number of $1$'s among the vectors $V'_1, V'_2, \dots, V'_M$, we have
\begin{align}
 \sum\nolimits_{l=1}^{Em} s'_l (M - s'_l)
 & = M \sum\nolimits_{l=1}^{Em} s'_l - \sum\nolimits_{l=1}^{Em} {s'_l}^2 \nonumber \\
 & \leq M \sum\nolimits_{l=1}^{Em} s'_l - \left(\sum\nolimits_{l=1}^{Em} s'_l\right)^2 / {Em} \nonumber \\
 & = M^2 e'_k - \frac{ M^2 {e'}_{k}^{2}}{Em} \label{equ:Johnsumdistbd}
\end{align}
Combining equations~\eqref{equ:Johnsumdistcolumn}, ~\eqref{equ:JohnsumHamming}, ~\eqref{equ:Johnsumsl} and~\eqref{equ:Johnsumdistbd}, and denote $\sum\nolimits_{k=1}^{n} e'_k = \bar{e} $ which is the average transform metric weight of the shifted codebook, we have
\begin{align}
 \sum\nolimits_{1 \leq i \leq j \leq M} d_{\TranE}(\Tran X'_i, \Tran X'_j) 
 & \leq \sum\nolimits_{k=1}^{n} \left( M^2 e'_k - \frac{ M^2 {e'}_{k}^{2}}{Em} \right) \nonumber \\
 & =  M^2  \sum\nolimits_{k=1}^{n}  e'_k - \frac{ M^2 }{Em}\sum\nolimits_{k=1}^{n}  {e'}_{k}^{2} \nonumber  \\
 & \leq M^2 \sum\nolimits_{k=1}^{n}  e'_k - \frac{ M^2 }{Em} \frac{ (\sum\nolimits_{k=1}^{n} e'_k)^2 }{n}   \nonumber \\
 & = M^2 \bar{e} - \frac{ M^2 {\bar{e}}^2}{Emn}. \label{equ:JohnsumdistII}
\end{align}
From equations~\eqref{equ:JohnsumdistI} and~\eqref{equ:JohnsumdistII}, we have
\begin{equation}
	 M(M-1)d/2 \leq M^2 \bar{e} - \frac{ M^2 {\bar{e}}^2}{Emn}. \nonumber
\end{equation}
Rearranging we have $\left(  d - 2\bar{e} + \frac{2\bar{e}^2}{Emn}  \right) M \leq d$. 
For $e \leq \frac{Emn}{2} \left( 1 - \sqrt{1-\frac{2d}{Emn}}  \right)$, note that the average weight is bounded from above by the radius, that is, $\bar{e} \leq e$, we also have $\bar{e} \leq \frac{Emn}{2} \left( 1 - \sqrt{1-\frac{2d}{Emn}}  \right)$. Hence,
\begin{align*}
	M 
	& \leq \frac{dEmn}{dEmn-2\bar{e}Emn+2\bar{e}^2} \\
	& = \frac{dEmn/2}{ (Emn/2 - \bar{e})^2 - (Emn/2 -d)Emn/2  }
\end{align*}
The denominator is positive, and it must be at least $1$ because it is an integer. Hence we have $J_{\TranE}(Cm \times n, d, e) \leq \frac{dEmn}{2}$ if $\frac{e}{Emn} < \frac{1}{2}\left( 1 - \sqrt{1-\frac{2d}{Emn}}  \right) $.
\hfill $\Box$

Using the Johnson-type bound, our Elias-Bassalygo-type bound is derived as follows.

We first prove that given a codebook $\calX$ of size $M$, for any $\eta$ there exists a transform metric ball $\calB_{\TranE}(\cdot,\eta)$ of radius $\eta$ containing at least $M \cdot \mathrm{Vol}\left( \calB_{\TranE}(\cdot,\eta)\right) /{2^{Cmn}}$ codewords.
 
Picking a transform metric ball $\calB_{\TranE}(\cdot,\eta)$ of radius $\eta$ around a random center. For each $X \in \calX$, let $\mathds{1}_X$ be an indicator variable of the event that $X \in \calB_{\TranE}(\cdot,\eta)$. The expected number of codewords from $\calX$ in the ball $\calB_{\TranE}(\cdot,\eta)$ is given by $E[\mathds{1}_X] = \Pr(\mathds{1}_X=1)=\mathrm{Vol}\left( \calB_{\TranE}(\cdot,\eta)\right)/{2^{Cmn}}$. Hence, 
\begin{align*}
E[\mbox{total number of codewords in }\calB_{\TranE}(\cdot,\eta) ] 
& = \sum E[\mathds{1}_X] \\
& = M \cdot \mathrm{Vol}\left( \calB_{\TranE}(\cdot,\eta)\right) /{2^{Cmn}} .
\end{align*}
There must be at least one ball achieving the expectation, hence there exists a transform metric ball $\calB_{\TranE}(\cdot,\eta)$ of radius $\eta$ containing at least $M \cdot \mathrm{Vol}\left( \calB_{\TranE}(\cdot,\eta)\right) /{2^{Cmn}}$ codewords.

Now set $\eta = \frac{Emn}{2}\left( 1 - \sqrt{1-\frac{2d}{Emn}}  \right) -1 $, by the Johnson-type bound in Lemma~\ref{thm:Johnson}, there can be no more than $\frac{dEmn}{2}$ codewords in the ball. Hence 
\begin{equation}
	M \cdot \mathrm{Vol}\left( \calB_{\TranE}(\cdot,\eta)\right) /{2^{Cmn}} \leq \frac{dEmn}{2}. \nonumber
\end{equation}
To obtain an upper bound on the codebook size $M$, we need to characterize a lower bound on the volume of the ball $\mathrm{Vol}\left( \calB_{\TranE}(\cdot,\eta)\right)$. Recall in~\eqref{equ:hammingtypestirling} in the proof of the Hamming-type bound in Theorem~\ref{thm:Hammingbd1} we have already bounded this quantity from below.
Note that the distance $d = 2pEmn+1$, 
\begin{align}
	\eta 
	& = \frac{Emn}{2}\left( 1 - \sqrt{1-\frac{2d}{Emn}}  \right) -1  \nonumber\\
	& \leq \frac{Emn}{2}\left( 1 - \sqrt{1-\frac{4pEmn+2}{Emn}}  \right)  \nonumber\\
	& \leq \frac{Emn}{2} ( 1 - \sqrt{1-4p}  ). \label{equ:eEmn}
\end{align}
If $p < \frac{C}{2Em} \left( 1 - \frac{C}{2Em}  \right)$, we have $\eta \leq Cn/2$. Recall in the proof of the Hamming-type bound, we can consider the matrices $Z$'s with $\eta \leq Cn/2$ $1$'s. All the matrices $\TranE Z$ are different for this specific set of $Z$. Hence the volume can be bounded from below by $\mathrm{Vol}\left( \calB_{\TranE}(\cdot,\eta)\right) \geq {Emn \choose \eta}$, which by Stirling's approximation is at least $2^{Emn \cdot H(\eta/{Emn}) - \log(Emn+1)}$.
In~\eqref{equ:eEmn} we bound $\eta$ from above, however we want to calculate $\eta/{Emn}$ more carefully now,
\begin{align*}
	\eta/{Emn} 
	& = \frac{1}{2}\left( 1 - \sqrt{1-\frac{2d}{Emn}}  \right) -1/{Emn} \\
	& = \frac{1}{2}\left( 1 - \sqrt{1-\frac{4pEmn+2}{Emn}}  \right) -1/{Emn} \\
	& = \frac{1}{2}\left( 1 - \sqrt{1-4p+\frac{2}{Emn}}  \right) -\frac{1}{Emn} .
\end{align*}

Hence,
\begin{align*}
	M
	& \leq  \frac{dEmn}{2} 2^{Cmn} /{\mathrm{Vol}\left( \calB_{\TranE}(\cdot,\eta)\right)} \\
	& \leq \frac{dEmn}{2}2^{Cmn \left( 1 -  \frac{E}{C} H(\eta/Emn) + \frac{\log(Emn+1)}{Cmn} \right)} \\
	& \leq \frac{dEmn}{2}2^{Cmn \left( 1 -  \frac{E}{C} H \left(\frac{1}{2}\left( 1 - \sqrt{1-4p+\frac{2}{Emn}}  \right) -\frac{1}{Emn} \right)  + \frac{\log(Emn+1)}{Cmn} \right) },
\end{align*}
which, asymptotically in $n$, gives the Elias-Bassalygo-type upper bound on the rate of any code as $1 -  \frac{E}{C} H \left(\frac{1 - \sqrt{1-4p}}{2} \right)+ \smallo\left(\frac{\log(Emn+1)}{Cmn}\right)$.

\section{Achievability} \label{sec:achieve}
In this section, we present some communication schemes over networks with worst-case bit-flip errors, and prove their corresponding achievable rates as presented in Theorems~\ref{thm:GV} and~\ref{thm:zyablovbd} in Section~\ref{sec:results}.
In Section~\ref{sec:achieve:GV},
we present several schemes motivated by the well-known Gilbert-Varshamov (GV) bound from classical coding theory~\cite{gilbert1952comparison,varshamov1957estimate} -- again, the challenge lies in deriving good {\it upper} bounds on the volume of spheres in the transform metric we define. 
Section~\ref{sec:achieve:GV:co} considers the {\it coherent} scenario, {\it i.e.}, when the linear coding coefficients in the network, or at least the transfer matrix $\Tran$ and the impulse response matrix $\TranE$, are known in advance to the receiver. 
We use this setting primarily for exposition, since the proof is somewhat simpler than the proof for the {\it non-coherent} setting, when no advance information about the topology of the network, the linear coding coefficients used, or $\Tran$ or $\TranE$ is known in advance to the receiver. 
In Section~\ref{sec:achieve:GV:nonco}, we are able to demonstrate that essentially the same rates are still achievable, albeit with an rate-loss that is asymptotically negligible in the block-length $\Bl$. 
Finally, Section~\ref{sec:achieve:zyablov} considers a concatenated version of the previously presented codes, so that the resulting codes' computational complexity scales polynomially in the block-length (albeit still exponentially in network parameters). The rate achieved by the concatenation scheme is characterized by a Zyablov-type lower bound. In Section~\ref{sec:achieve:zyablov:Dec}, a {\it Generalized Minimum Distance Decoding} scheme is provided, which is able to correct up to half of the minimum distance of the concatenated codes.
\vspace{-0.5cm}
\subsection{Gilbert-Varshamov-type bounds}\label{sec:achieve:GV}
\subsubsection{Coherent GV-type network codes}\label{sec:achieve:GV:co}
We first discuss the case when the network transfer matrix $\Tran$ and impulse response matrix $\TranE$ are known in advance.

{\it Codebook design:} Initialize the set $\GVset$ as the set of all binary $\Cp \lfs \times \Bl$ matrices.
Choose a uniformly random $\Cp \lfs\times \Bl$ binary matrix $X$ as the first codeword. Eliminate from $\GVset$ all matrices in the radius-$2p\NoE\lfs\Bl$ ball (in the transform metric)
$\calB_{\TranE}(\Tran X,2p\NoE\lfs\Bl)$.
Then choose a matrix $Y'$ uniformly at random in the remaining set and choose $X'=\Tran^{-1}Y'$ as the second codeword. Now, further eliminate all matrices in the radius-$2p\NoE\lfs\Bl$ ball
$\calB_{\TranE}(\Tran X',2p\NoE\lfs\Bl)$ from $\GVset$, choose a random $Y'$ from the remaining set, and choose the third codeword $X''$ as $X''=\Tran^{-1}Y''$.
Repeat this procedure until the set $\GVset$ is empty.

{\it Decoder:} The receiver uses a minimum distance decoder with the transform metric, that is, the decoder picks the codeword $X$ which minimizes the transform metric distance $d_{\TranE}(\Tran X,Y)$ between $\Tran X$ and the received matrix $Y$.

To prove Theorem~\ref{thm:GV}.\ref{thm:GVbdco}, we need an upper bound on $\calB_{\TranE}(\Tran X,2p\NoE\lfs\Bl)$ (rather than a lower bound on $\calB_{\TranE}(\Tran X,p\NoE\lfs\Bl)$ as in Section~\ref{sec:converse:Hamming}).
Recall that $Y=\Tran X+ \TranE Z$
The number of different $Y$, or equivalently, different $\TranE Z$, can be bounded from above by the number of different $Z$. This equals $\displaystyle\sum\limits_{i=0}^{2p\NoE \lfs \Bl} {{\NoE \lfs \Bl}\choose{i}}$. The dominant term this summation is when $i$ equals $2p\NoE \lfs \Bl$. Hence the summation can be bounded from above by $(2p\NoE \lfs \Bl+1){{\NoE \lfs \Bl}\choose{2p\NoE \lfs \Bl}}$. By Stirling's approximation~\cite{CoverT:91} we have that
\begin{equation} \label{eq:GVcostirling}
 |\calB_{\TranE}(\Tran X,2p\NoE\lfs\Bl)| \leq (2p\NoE \lfs \Bl+1)2^{H(2p)\NoE\lfs\Bl}. \nonumber
\end{equation}

Thus a lower bound on the size of the codebook for coherent GV-type
\begin{equation*}
    \frac{2^{\Cp \lfs \Bl}}{(2p\NoE \lfs \Bl+1)2^{H(2p)\NoE \lfs \Bl}} = 2^{ \left( 1-\frac{\NoE}{\Cp}H(2p)-\frac{\log(2p\NoE \lfs \Bl+1)}{C m \Bl} \right)\Cp \lfs \Bl},
\end{equation*}
which, asymptotically in $\Bl$, gives the rate of coherent GV-type bound network codes as $1-\frac{\NoE}{\Cp}H(2p) - \smallo\left(\frac{\log(2p\NoE \lfs \Bl+1)}{Cm\Bl}\right)$.

\subsubsection{Non-coherent GV-type network codes}\label{sec:achieve:GV:nonco}
The assumption that $\Tran$ and $\TranE$ are known in advance to the receiver is often unrealistic, since the random linear coding coefficients in the network are usually chosen on the fly.
Hence we now consider the non-coherent setting, wherein $\Tran$ and $\TranE$ are not known in advance. We demonstrate that despite this lack of information the same rates as in Theorem~\ref{thm:GV}.\ref{thm:GVbdco} are achievable in the non-coherent setting.

The number of all possible $\TranE$ is at most by $2^{\Cp\NoE\lfs}$ since $\TranE$ is a ${\Cp\times \NoE}$ matrix $ \Fbm$ -- the crucial observation is that this number is independent of the block-length $\Bl$.
Hence in the non-coherent GV setting, we consider {\it all} possible values of $\TranE$, and hence $\Tran$, since it comprises of a specific subset of $\Cp$ columns of $\TranE$.

{\it Codebook design:} Initialize the set $\GVset$ as the set of all binary $\Cp \lfs \times \Bl$ matrices.
Choose a uniformly random $\Cp \lfs\times \Bl$ binary matrix $X$ as the first codeword. For each
$\Cp\times \NoE$ matrix $\TranE$ (over the field $\Fbm$),
eliminate from $\GVset$ all matrices in the radius-$2p\NoE\lfs\Bl$ ball (in the transform metric)
$\calB_{\TranE}(\Tran X,2p\NoE\lfs\Bl)$.
Then choose a matrix $Y'$ uniformly at random in the remaining set and choose $X'=\Tran^{-1}Y'$ as the second codeword. Now, further eliminate all matrices in the radius-$2p\NoE\lfs\Bl$ ball
$\calB_{\TranE}(\Tran X',2p\NoE\lfs\Bl)$ from $\GVset$, choose a random $Y'$ from the remaining set, and choose the third codeword $X''$ as $X''=\Tran^{-1}Y''$.
Repeat this procedure until the set $\GVset$ is empty.

{\it Decoder:} The receiver uses a minimum distance decoder with the transform metric, that is, the decoder picks the codeword $X$ which minimizes the transform metric distance $d_{\TranE}(\Tran X,Y)$ between $\Tran X$ and the received matrix $Y$ for all possible $\TranE$.

The crucial difference with the proof of Theorem~\ref{thm:GV}.\ref{thm:GVbdco} is in the process of choosing codewords -- at each stage of the codeword elimination process, at most $2^{\Cp\NoE\lfs}|\calB_{\TranE}(\Tran X',2p\NoE\lfs\Bl)|$ potential codewords are eliminated (rather than $|\calB_{\TranE}(\Tran X',2p\NoE\lfs\Bl)|$ potential codewords as in Theorem~\ref{thm:GV}.\ref{thm:GVbdco}). Hence the number of potential codewords that can be chosen in the codebook is at least
$$\frac{2^{\Cp \lfs \Bl}}{2^{\Cp\NoE\lfs}(2p\NoE \lfs \Bl+1)2^{H(2p)\NoE \lfs \Bl}}$$
which equals$$
2^{\left(1-\frac{\NoE}{\Cp}H(2p)-\frac{\log(2p\NoE \lfs \Bl+1)+\NoE}{\Bl}\right)\Cp \lfs \Bl}.
$$ As can be verified, asymptotically in $\Bl$ this leads to the same rate of $1-\frac{\NoE}{\Cp}H(2p) - \smallo\left( \frac{\log(2p\NoE \lfs \Bl+1)+\NoE}{\Bl} \right)$ as in Theorem~\ref{thm:GV}.\ref{thm:GVbdco}.

{\it Note:} We show in the following Section~\ref{sec:achieve:GV:linear} that random linear codes achieve the GV-type bound with high probability, which reduce the encoding complexity.

\subsubsection{Linear GV-Type Bound}\label{sec:achieve:GV:linear}
Similar to Varshamov's linear construction~\cite{varshamov1957estimate} in classical coding theory, we show that for our worst-case binary-error network channel, random linear codes achieve the GV-type bound with high probability.

Let $G \in \F_2^{km \times n}$ be the generator matrix of a random linear code, where each entry of $G$ is chosen uniformly and independently at random from $\F_q$. Let $ M \in \F_2^{Cm \times (k-C)m} \backslash \mathbf{0}$ and $\bar{M} = [I \; M]$ be a $Cm \times km$ matrix by sticking a $Cm \times Cm$ identity matrix $I$ in front of $M$. (The parameter $k$ here should be sufficiently large so that $k -C> 0$.) We need to show that for any matrix $M \in \F_2^{Cm \times (k-C)m} \backslash \mathbf{0}$, $d_{\TranE}(\bar{M}G, \mathbf{0}) \geq d$ with high probability, where $d=2pEmn+1$ is the minimum distance we require for the codebook.\footnote{Note that in $q$-ary alphabet, $ M \in \F_q^{C \times (k-C)} \backslash \mathbf{0}$. Hence, the number of all possible $M$ is $2^{C(k-C)m}$.}

Note that for any fixed matrix $ M \in \F_2^{Cm \times (k-C)m} \backslash \mathbf{0}$, by choosing $G$ uniformly at random, $\bar{M}G$ is a uniformly random matrix from $\F_2^{Cm \times n}$.  Hence, the probability over the choice of $G$ of the code being ``bad" can be bounded from above in the following way,
\begin{align*}
\Pr(d_{\TranE}(\bar{M}G, \mathbf{0}) < d) 
& = \frac{|\calB_{\TranE}(\mathbf{0},d-1)|}{2^{Cmn}} \\
& = \frac{|\calB_{\TranE}(\mathbf{0},2p\NoE\lfs\Bl)|}{2^{Cmn}} \\
& \leq \frac{(2p\NoE \lfs \Bl+1)2^{H(2p)\NoE\lfs\Bl}}{2^{Cmn}},
\end{align*}
where the last inequality is as in~\eqref{eq:GVcostirling} in the proof of Theorem~\ref{thm:GV}.\ref{thm:GVbdco} in Section~\ref{sec:achieve:GV:co}. By the union bound,
\begin{align*}
\Pr(\exists M, d_{\TranE}(MG, \mathbf{0}) < d)
& \leq 2^{Cm(k-C)} \cdot \frac{(2p\NoE \lfs \Bl+1)2^{H(2p)\NoE\lfs\Bl}}{2^{Cmn}} \\
& = (2pEmn+1)2^{-\varepsilon Cmn},
\end{align*}
if we choose $k = \left( 1-\frac{E}{C}H(2p)-\varepsilon \right)n +C$. Since $(2pEmn+1)2^{-\varepsilon Cmn} \ll 1$ for large enough $n$, we have shown that there exists a linear code with minimum distance $2pEmn+1$ and rate at least $ 1-\frac{E}{C}H(2p)-\varepsilon$.


{\it Note:} The advantage of this Varshamov-type construction is that the encoding complexity is $\bigO(n^2Cm^2)$, though the decoding complexity is still $\Omega(e^n)$. To deal with the high decoding complexity, we now suggest a concatenated construction in the following Section~\ref{sec:achieve:zyablov} so that their encoding and decoding complexity grows only polynomial in the block-length (albeit still exponentially in network parameters). 
\vspace{-0.5cm}
\subsection{Concatenated Codes and Zyablov-Type Bound} \label{sec:achieve:zyablov}
The codes which achieve the Gilbert-Varshamov-type bound in Section~\ref{sec:achieve:GV} take running time $2^{\bigO{(n)}}$. This section provides an code concatenation strategy using the GV-type code from Section~\ref{sec:achieve:GV:co} as the {\it inner code} and a Reed-Solomon code as the {\it outer code}. This type of {\it concatenated network codes} have encoding/decoding complexity that is polynomial in the block length $n$ (albeit still exponentially in the network parameter $C$ and the coding parameter $m$). 
Also, a Zyablov-type lower bound stated in Theorem~\ref{thm:zyablovbd} is proven, which characterizes the rate achieved by the concatenated network codes.

\subsubsection{Code Concatenation Construction} \label{sec:achieve:zyablov:Enc}
Consider the outer code and the inner code as follows,
\begin{align*}
	\Cout 
	& : \big[ \F_{2^{Cm \times \Rin \log{n} } } \big]^{ \Rout \frac{n}{\log{n}}} \rightarrow \big[ \F_{2^{Cm \times \Rin \log{n} } } \big]^{ \frac{n}{\log{n}}} , \\
	\Cin
	& : \big[ \F_2 \big]^{Cm \times \Rin \log{n} } \rightarrow \big[ \F_2 \big]^{Cm \times  \log{n} },
\end{align*}
where $\Rout$ and $\Rin$ are the corresponding rates of the outer and inner codes to be characterized later.
The concatenated code is denoted by $\Ccon = \Cout \circ \Cin $, and conducts the following steps.
\begin{itemize}
	\item Firstly, the encoder breaks the messages from $\F_2^{Cm \times \Rout \Rin n}$ into $\Rout \frac{n}{\log{n}}$ such many chunks with size $Cm \times \Rin \log{n}$, and treats each chunk as an element from the large field $\F_{2^{Cm \times \Rin \log{n} } }$. The field size is much larger than the block length $\frac{n}{\log{n}}$, hence one can take an $\Big[ \frac{n}{\log{n}}, \Rout \frac{n}{\log{n}},$ $\dout \Big]_{2^{Cm \times \Rin \log{n} } }$ Reed-Solomon code as the outer code. Therefore, the minimum distance of the outer code is $\dout = (1-\Rout) \frac{n}{\log{n}} +1.$
	The outer code converts the messages into codewords of length $\frac{n}{\log{n}}$ over the large alphabet $\F_{2^{Cm \times \Rin \log{n} } }$.
	
	\item Secondly, the encoder takes the output codewords from the outer code and converts the symbols from the field $\F_{2^{Cm \times \Rin \log{n} } }$ into binary matrices of size $Cm \times \Rin \log{n} $. For the inner code, the encoder takes the block binary matrices from $ \big[ \F_2 \big]^{Cm \times \Rin \log{n} } $ as messages, then uses the same codebook design for the GV-type bound as in Section~\ref{sec:achieve:GV:co}. Hence the minimum distance $\din$ and the rate $\Rin$ of the inner code satisfy $\Rin = 1 - \frac{E}{C} H \left( \frac{\din}{Em \log{n}} \right),$
	which gives that $\din = H^{-1}\left( \frac{C}{E}(1-\Rin) \right) Em \log{n} $. This completes the whole concatenated coding process and outputs codewords from $\F_2^{Cm \times n}$.
\end{itemize}

\subsubsection{Zyablov-Type Bound} \label{sec:achieve:zyablov:proof}

The Reed-Solomon outer code has minimum distance $\dout = (1-\Rout) \frac{n}{\log{n}} +1$.
The GV-type inner code has minimum distance $\din = H^{-1}\left( \frac{C}{E}(1-\Rin) \right) Em \log{n}$.
The overall distance $D$ of the code $\Ccon$ satisfies 
$	D \geq \dout \cdot \din \geq  (1-\Rout) H^{-1}\left( \frac{C}{E}(1-\Rin) \right) Emn.  $

Take $(1-\Rout) H^{-1}\left( \frac{C}{E}(1-\Rin) \right) Emn = 2pEmn$, we have $\Rout = 1 - \frac{2p}{H^{-1}\left( \frac{C}{E}(1-\Rin) \right)}$. The overall rate of the concatenated code is $R = \Rout \cdot \Rin$, replace $\Rin$ by a adjustable variable $r$, optimized over the choice of $r$, the rate of the concatenated code satisfies
\begin{equation}
	R \geq \max_{0<r<1-\frac{E}{C}H(2p)} r \cdot \left( 1 - \frac{2p}{H^{-1}\left( \frac{C}{E}(1-r) \right)} \right), \nonumber
\end{equation}
where the constraint $r<1-\frac{E}{C}H(2p)$ is necessary to guarantee that $R>0$.

\subsubsection{Generalized Minimum Distance Decoding} \label{sec:achieve:zyablov:Dec}
A natural decoding algorithm is to reverse the encoding process as described in Section~\ref{sec:achieve:zyablov:Enc}. Briefly, the algorithm uses the inner code to decode each chunk with possibly wrongly decoded chunks, then uses the outer code to correct the wrongly decoded chunks. 
Denote the input vector to the decoder as $Y = (Y_1, Y_2, \dots ,Y_{n/\log{n}} ) \in \big[ (\F_{2})^{Cm \times \log{n}} \big]^{n/\log{n}}$.
The natural decoding algorithm is described as follows.

\noindent {\it Natural decoding algorithm:}
\begin{description}
\item[Step1:] Decode each $Y_i$ to $V_i \in \F_{2^{Cm \times \Rin \log{n} } }$ such that $V_i$ minimizes $d_{\TranE}\left(\Cin(V_i), Y_i\right)$.
\item[Step2:] Decode $V = (V_1, V_2, \dots, V_{n/\log{n}})$ using decoding algorithms for the RS outer code.
\end{description}
It can be easily shown that the natural decoding algorithm can correct up to $ ( \dout \cdot \din )/4$ errors. Briefly, the outer code fails only if the number of wrongly decoded inner chunks is greater than $\dout /2$. An inner chunk is decoded wrongly only when there are more than $\din/2$ errors.

To improve the decodability to correct up to half the minimum distance $ ( \dout \cdot \din )/2$, we develop the algorithm below mimicking the {\it generalized minimum distance decoding}~\cite{forney1966generalized} for classical concatenated codes. 

\noindent {\it Generalized minimum distance (GMD) decoding algorithm:}
\begin{description}
\item[Step1:] Decode each $Y_i$ to $V_i \in \F_{2^{Cm \times \Rin \log{n} } }$ such that $V_i$ minimizes $d_{\TranE}\left(\Cin(V_i), Y_i\right)$. Let $\omega_i = \min \big( d_{\TranE}\left(\Cin(V_i), Y_i\right), \din /2 \big)$.
\item[Step2:] With probability $\frac{2 \omega_i}{\din}$, set $V_i' = ?$ to be an erasure; otherwise, set $V_i' = V_i$.
\item[Step3:] Decode $V' = (V_1', V_2', \dots, V_{n/\log{n}}')$ with both errors and erasures using decoding algorithms for the RS outer code.
\end{description}
Denote the number of errors by $e$ and number of erasures by $s$, an RS code with minimum distance $\dout$ can decode correctly if $2e+s < \dout$. The following Lemma shows that in expectation it is indeed the case if the total number of errors is less than $( \dout \cdot \din )/2$.
\begin{lemma} \label{thm:GMDran}
Let $W = (W_1, W_2, \dots, W_{n/\log{n}})$ be the codeword sent, suppose $d_{\TranE}(W, Y) < ( \dout \cdot \din )/2$ holds. If $V'$ has $e$ errors and $s$ erasures compared to $W$, then $E[2e+s] < \dout$.  
\end{lemma}
\noindent {\it Proof:}
For $1 \leq i \leq n/\log{n}$, let $\delta_i = d_{\TranE}(W_i, Y_i)$, then 
\begin{equation}
	\sum_{i=1}^{n/\log{n}} \delta_i < \frac{\dout \cdot \din }{2}. \label{eq:sumdelta}
\end{equation}
Define two indicator random variables $\mathds{1}_{i}^{err}$ and $\mathds{1}_i^{ers}$ for the event of an error and an erasure at $V_i'$ respectively. In the following, we show through case analysis that
\begin{equation}
E[2 \cdot \mathds{1}_{i}^{err} + \mathds{1}_i^{ers}] \leq \frac{2 \delta_i}{\din}. \label{eq:expindi}
\end{equation}
\noindent {\bf Case 1 ($W_i = \Cin (V_i)$).} For the erasure event, we have $E[\mathds{1}_i^{ers}] = \Pr(\mathds{1}_i^{ers} = 1) = \frac{2 \omega_i}{\din}$. For the error event, if $V_i' = ?$ is an erasure, then $\mathds{1}_{i}^{err} = 0$; otherwise $W_i = \Cin (V_i) = \Cin(V_i')$, which means there is no error $\mathds{1}_{i}^{err} = 0$. By the definition of $\omega_i$, we have $\omega_i \leq d_{\TranE}\left(\Cin(V_i), Y_i\right) = d_{\TranE}\left(W_i, Y_i\right) = \delta_i$.
Hence, in this case $E[2 \cdot \mathds{1}_{i}^{err} + \mathds{1}_i^{ers}] = \frac{2 \omega_i}{\din} \leq \frac{2 \delta_i}{\din}$.

\noindent {\bf Case 2 ($W_i \neq \Cin (V_i)$).} In this case, still we have $E[\mathds{1}_i^{ers}] = \frac{2 \omega_i}{\din}$. When $V_i'$ is not an erasure we have $W_i \neq \Cin (V_i')$, which means that $V_i'$ has an error. Hence, $E[\mathds{1}_i^{err}] = 1 - \Pr(\mathds{1}_i^{ers} = 1) = 1 - \frac{2 \omega_i}{\din}$ and $E[2 \cdot \mathds{1}_{i}^{err} + \mathds{1}_i^{ers}] = 2 - \frac{2 \omega_i}{\din}$. In the following, we show that $\omega_i + \delta_i \geq \din$ through case analysis.
\begin{itemize}
\item Case 2.1 ($\omega_i = d_{\TranE}\left(\Cin(V_i), Y_i\right) <  \din /2$). In this case, $\omega_i + \delta_i = d_{\TranE}\left(\Cin(V_i), Y_i\right)+  d_{\TranE}(W_i, Y_i) \geq  d_{\TranE}(\Cin(V_i), W_i) \geq \din$, where the first inequality is by triangle inequality and the second inequality follows by the minimum distance of the codebook since $W_i \neq \Cin (V_i)$ are two different codewords.
\item Case 2.2 ($\omega_i = \din /2 \leq  d_{\TranE}\left(\Cin(V_i), Y_i\right)$ ). In this case, $\delta_i = d_{\TranE}(W_i, Y_i) \geq d_{\TranE}\left(\Cin(V_i), Y_i\right) \geq \din /2 $, where the first inequality is by the fact that we decode $Y_i$ to $V_i$ which minimize the transform metric distance. Hence, $\omega_i + \delta_i \geq \din$.
\end{itemize}
Hence for Case 2, $E[2 \cdot \mathds{1}_{i}^{err} + \mathds{1}_i^{ers}] = 2 - \frac{2 \omega_i}{\din} \geq \frac{2 \delta_i}{\din}$.
Hence we have shown \eqref{eq:expindi}, and combining with \eqref{eq:sumdelta} we have 
\begin{align*}
E[2e+s]
& = E[ \sum\nolimits_{i=1}^{n\log{n}}  2 \cdot \mathds{1}_{i}^{err} + \mathds{1}_i^{ers}  ] \\
& = \sum\nolimits_{i=1}^{n\log{n}} E[2 \cdot \mathds{1}_{i}^{err} + \mathds{1}_i^{ers}] \\\
& \leq \sum\nolimits_{i=1}^{n\log{n}} \frac{2 \delta_i}{\din} \\
& < \frac{2 }{\din} \cdot \frac{\dout \cdot \din }{2} \\
& = \dout.
\end{align*}
\hfill $\Box$

\vspace{-1cm}
\section{Conclusion}
In this work we investigate upper and lower bounds for the performance of end-to-end error-correcting codes for worst-case binary errors. This model is appropriate for highly dynamic wireless networks, wherein the noise-levels on individual links might be hard to accurately estimate. We demonstrate significantly better performance for our proposed schemes, compared to prior benchmark schemes. 

\bibliographystyle{IEEEtran}
\bibliography{IEEEabrv,database}

\end{document}